\documentclass[12pt]{iopart}
\usepackage[utf8]{inputenc}   
\usepackage[english]{babel}   
\usepackage{blindtext}
\usepackage{iopams}
\usepackage{setstack}
\usepackage{graphicx}
\usepackage{color}
\usepackage{cite}

\newcommand{\cblue}{\color{black}}

\begin{document}
\title[Dark solitons in the unitary Bose gas]
{Dark solitons in the unitary Bose gas}
\author{M. Calzavara and L. Salasnich}
\address{Dipartimento di Fisica e Astronomia "Galileo Galilei", 
Universita di Padova, \\ Via Marzolo 8, 35131, Padova, Italy}


\begin{abstract}
We study the dilute and ultracold unitary Bose gas, 
characterized by a universal equation of state 
due to the diverging s-wave scattering length, under a 
{\cblue transverse} harmonic {\cblue confinement}. 
From the hydrodynamic equations of superfluids we 
derive an effective one-dimensional nonpolynomial Schr\"odinger 
equation (1D NPSE) for the axial {\cblue wavefunction} which, however, 
takes also into account the transverse {\cblue wavefunction}. 
Finally, by solving the 1D NPSE we obtain meaningful analytical 
formulas for the dark (gray and black) solitons of the bosonic system. 
\end{abstract}



\section{Introduction}

The possibility of tuning the interaction between atoms near Feshbach 
resonances \cite{Inouye1998} almost at will, {\cblue by precision control of external 
magnetic fields}, is allowing the experimenters to explore the physics of ultracold gases in a wide range of 
values of the coupling constant {\cblue characterized by the s-wave scattering length.  
The possibility to control the scattering length via Feshbach resonances makes 
ultracold atomic gases an excellent setting for studies of strongly correlated behavior.} 
In particular it made possible to observe first a Fermi gas 
\cite{OHara2002, Regal2003, Strecker2003,Jochim2003, Gehm2003} and then 
also a Bose gas at unitarity \cite{Li2012, Rem2013, Makotyn2014, Fletcher2013},
that is the limit in which the s-wave scattering length is infinitely large. 
{\cblue In this so-called unitary regime, the interactions are as strong as allowed by quantum mechanics, 
and the physics cannot explicitly depend on the scattering length, 
leading to the possibility of new types of universal behavior.} 
In the case of a Bose gas, the presence of three body recombination 
processes \cite{Roberts2000} complicates the experimental realization 
of the unitary limit, which can be held for a limited amount of time. 
Nevertheless, the peculiar properties of a unitary gas, which behaves in a universal manner and 
exhibits scale invariance, encourage us to study the physics of such systems more in depth. 

In this paper we {\cblue investigate solitonic configurations} 
of the {\cblue unitary Bose gas}. More specifically, we {\cblue analyze} 
{\cblue axial and transverse density profiles} 
of dark solitons in bosonic alkali-metal atoms at unitarity 
{\cblue under the action of a transverse harmonic confinement}. 
A dark soliton is a self-bound solitary wave, whose 
existence is made possible by the {\cblue interplay between 
the repulsive interaction and the presence of a phase gradient}. 
A dark soliton {\cblue is characterized by} a local decrease in density with 
respect to a uniform background. {\cblue The depth of the dark soliton 
crucially depends on the phase angle of the complex 
field which describes the Bose gas}.
 
The realizability of {\cblue black and gray} solitons in ultracold gases is well documented 
in the weak-coupling limit \cite{Burger1999, Denschlag2000} and 
in a unitary Fermi gas \cite{Ku2016}. Moreover, the properties 
of {\cblue black} solitons have been theoretically investigated for superfluid fermions 
also in the BCS-BEC crossover \cite{Antezza2007,Cetoli2013,Lombardi2017,Alphen2018}. 
It is then a natural question to ask, whether {\cblue black and gray} solitons 
can be realized also in the case {\cblue of} Bose gases at unitarity, and what are their specific properties. 
{\cblue It is important to stress that dark (black and gray) solitons of the unitary Bose gas 
{\cblue have} not yet been observed experimentally nor theoretically analyzed. 
Our theoretical paper can help experimental groups because we are giving analytical 
and semi-analytical formulas for density profiles, widths, and velocities of these 
dark solitons. All these quantities are universal because, contrary to the dark 
solitons of weakly-interacting bosonic gases, they do not depend on the s-wave scattering length.}  
 
We first derive the equations which govern the dynamics of 
these systems. The resulting equation {\cblue is} a nonlinear 
Schr\"{o}dinger equation for a complex order parameter 
$\psi({\bi r},t)=|\psi({\bi r},t)|e^{i\alpha({\bi r},t)}$ 
\cite{Adhikari2008}, where $n({\bi r},t)=|\psi({\bi r},t)|^2$ is the 
local number density of the bosonic system and 
${\bi v}({\bi r},t)=(\hbar/m)\nabla \alpha({\bi r},t)$ is the local velocity 
with $\alpha({\bi r},t)$ the phase angle of the complex field. 
Then, we find the solutions which are in the form of dark solitons,  
{\cblue specifically as stationary objects in a moving frame}. 

Mimicking the experimental practice of confining ultracold gases in the 
minimum of a potential, we study the case in which the Bose gas 
is placed in a harmonic potential with cylindrical symmetry, which 
is an approximation to a more realistic cigar shaped 
potential \cite{Burger1999, Ku2016}.
A Gaussian variational approach, whose effectiveness in similar 
cases has already been proved 
\cite{Salasnich2002, Salasnich2004, Salasnich2005, Salasnich2006, 
Salasnich2007_1, Salasnich2007_2}, is then deployed to obtain an 
effective one-dimensional nonpolynomial Schr\"{o}dinger equation (1D NPSE) 
for the axial wavefunction. 
{\cblue The relevance of using the 1D NPSE in the study of solitons 
is due to the fact that the obtained solutions are a reliable 
generalization of familiar strictly one-dimensional results. 
These generalized solutions take properly into account that 
the transverse width of the cigar-shaped bosonic cloud depends 
on the axial coordinate}.

In particular, we focus on the case of weak {\cblue transverse} confinement, 
in which more naive attempts to reduce the dimensionality of the problem fail.
We integrate analytically the 1D NPSE thanks to the presence of a 
constant of motion, leaving behind a solution in terms of an integral, 
which we finally evaluate by means of numerical techniques. 
We are then able to find {\cblue the} axial density profile, transverse width and phase of 
the bosonic system. Moreover, we find the relation between 
density at the minimum and velocity of the soliton. 
While doing so, we also find that the theory actually depends on just 
one free parameter. Finally, we comment on the limits of 
the weak-coupling approximation.

\section{From Euler to Schr\"{o}dinger}

Euler equations describe the dynamics of a non-viscous and irrotational fluid, 
such as a superfluid Bose gas at zero temperature \cite{pitaevskii2003}. 
In presence of an external potential $U(\bi{r})$, these equations are given by
\begin{eqnarray}
\frac{\partial n}{\partial t} + \nabla \cdot (n\bi{v}) = 0 \label{eq:eul1}\\
m\frac{\partial \bi{v}}{\partial t} + \nabla (U(\bi{r}) + 
An^{\frac{2}{3}} +\frac{1}{2}m\bi{v}^2)=0 \label{eq:eul2}.
\end{eqnarray}
The equation of state of the unitary Bose gas at zero temperature 
is assumed to be
\begin{equation}
P(n)= \frac{2}{5} \, A \, n^{\frac{5}{3}},
\end{equation}
where $n$ is the density of particles and $A=u\frac{\hbar^2}{m}$, 
with $u$ a universal and adimensional (positive) coefficient. 
In fact, we expect this to be the case in the unitary limit 
$a_s \rightarrow \infty$ because of dimensional analysis, 
taking into account that the only important length scale 
is the mean distance between atoms $\approx n^{-\frac{1}{3}}$.
The value of $u=\xi (6\pi^2)^{\frac{2}{3}}\frac{5}{6}$ has been derived 
theoretically in many ways 
\cite{Cowell2002,Song2009,Lee2010,Diederix2011,Rossi2014} and, depending on 
the procedure, values in the range $\xi \approx 0.4 \div 1.75$ have been 
reported.

We shall now derive a nonlinear Schr\"{o}dinger equation (NLSE) with a 
${4}/{3}$ nonlinearity exponent \cite{Adhikari2008}, starting 
from equations \eref{eq:eul1}-\eref{eq:eul2} and adding a term, 
which will account for quantum effects.
This is done via the mapping 
\begin{eqnarray}
n(\bi{r},t) = | \psi(\bi{r},t) |^2 \label{eq:map1}
\\
\bi{v}(\bi{r},t) = \frac{\hbar}{m} \nabla \alpha(\bi{r},t) , 
\label{eq:map2}
\end{eqnarray}
{\cblue where $\alpha({\bi r},t)$ is a scalar field. 
Equation (\ref{eq:map2}) is indeed fully justified by the fact 
the fluid is irrotational, i.e. $\nabla \wedge \bi{v}= {\bf 0}$.} 

If we substitute \eref{eq:map1}-\eref{eq:map2} into 
\eref{eq:eul1}-\eref{eq:eul2} we obtain
\begin{eqnarray}
\frac{\partial |\psi |}{\partial t} + \frac{\hbar}{2m}(\nabla^2 \alpha)
|\psi |+ \frac{\hbar}{m} \nabla \alpha \cdot \nabla |\psi |= 0  
\label{eq:sost1}\\
\hbar \nabla \frac{\partial \alpha}{\partial t} + \nabla (U(\bi{r}) + 
An^{\frac{2}{3}} +\frac{\hbar^2}{2m}|\nabla \alpha|^2)=0 \label{eq:sost2}.
\end{eqnarray}
Let us integrate equation \eref{eq:sost2} and multiply both sides by $|\psi|$
\begin{equation}
-|\psi |\hbar \frac{\partial \alpha}{\partial t} = 
|\psi |(U(\bi{r}) + An^{\frac{2}{3}} +\frac{\hbar^2}{2m}|\nabla \alpha|^2 + C).
\label{eq:sost_int}
\end{equation}
The integration constant $C$ appeared, but we shall see that it is not 
going to affect the dynamics. 
In order to get the Schr\"{o}dinger equation we now add the term 
$-\frac{\hbar^2}{2m} \nabla^2 | \psi({\bi r},t) |$ 
on the right hand side of equation \eref{eq:sost_int}. 
This is chosen to obtain the same kinetic term which appears in 
the weak-coupling limit $P \propto n$. 
In that case we would obtain the Gross-Pitaevskii equation, whose validity 
for weak coupling is well established \cite{Gross1961,Pitaevskii1961}.

We can express the system of real equations as a single complex equation 
by adding equation \eref{eq:sost1} multiplied by $i\hbar$ to equation 
\eref{eq:sost_int}. Then we multiply every term by $e^{i\alpha}$, to get
\begin{eqnarray}
\fl \left[ i \hbar \frac{\partial | \psi |}{\partial t} 
- \hbar | \psi | \frac{\partial \alpha}{\partial t} \right] e^{i\alpha} 
&=&  \left[ - \frac{\hbar^2}{2m}(-| \psi | | \nabla \alpha|^2 
+ \nabla^2 | \psi | + i\nabla^2 \alpha | \psi | + 
2i \nabla \alpha \cdot \nabla | \psi |) \right]e^{i\alpha}
\nonumber
\\ 
&+& \left[(U(\bi{r}) + An^{\frac{2}{3}} + C) | \psi | \right] e^{i\alpha} 
\label{eq:cpx}.
\end{eqnarray}
Finally, by defining
\begin{equation}
\psi({\bi r},t) = | \psi({\bi r},t) | \, e^{i\alpha({\bi r},t)}
\label{fattoio} 
\end{equation}
last equation can be written as a NLSE
\begin{equation}
i\hbar \frac{\partial \psi}{\partial t} = -\frac{\hbar^2}{2m} \nabla^2 \psi 
+ (U(\bi{r}) + C) \psi + A | \psi |^{\frac{4}{3}} \psi.
\label{eq:nlse}
\end{equation}
$C$ results in a potential energy offset. It can be removed by substituting 
$\psi \rightarrow e^{-\frac{i}{\hbar}Ct}\psi$. 
{\cblue It is important to stress that equation (\ref{fattoio}) implies that 
the scalar field $\alpha({\bi r},t)$ is an angle and the circulation 
of the velocity field ${\bi v}({\bi r},t)=(\hbar/m)\nabla \alpha({\bi r},t)$ 
is quantized.} 

\section{Nonpolynomial Schr\"{o}dinger equation}

\Eref{eq:nlse} can also be derived minimizing the action 
functional $S=\int dt\ L$
\begin{equation}
\fl S[\psi,\psi^{\star}] =\int_{t_{1}}^{t_{2}} dt \int d^3\bi{r}\ \frac{i\hbar}{2} 
\left( \psi^{\star} \frac{\partial \psi}{\partial t} - \psi \frac{\partial 
\psi^{\star}}{\partial t} \right) - \frac{\hbar^2}{2m} |\nabla \psi|^2 
- U(\bi{r})| \psi |^2 - A |\psi|^{\frac{10}{3}}
\label{eq:az}
\end{equation}
with the constraint
\begin{equation}
\int d^3\bi{r}\ | \psi(\bi{r},t) |^2 = N.
\label{eq:const}
\end{equation}

Let us suppose that the confinement potential is harmonic, with cylindrical 
symmetry along the $z$ axis,
\begin{equation}
U(\bi{r}) = \frac{1}{2}m \omega_{\perp}^2 (x^2 + y^2)
\end{equation}
and consider the following variational ansatz for $\psi$ 
\begin{equation}
\psi(\bi{r},t) = \phi(z,t)\frac{exp \left(-\frac{x^2+y^2}{2 \sigma^2(z,t)} 
\right)}{\sqrt{\pi \sigma^2(z,t)}}, 
\label{eq:ansz}
\end{equation}
where the variational parameters are {\cblue the axial wavefunction} 
$\phi(z,t)$ and {\cblue the width} $\sigma(z,t)$ {\cblue of the 
transverse Gaussian wavefunction. The choice of a Gaussian function 
in the $(x,y)$ plane is justified by the presence of a harmonic 
potential in the $(x,y)$ plane. Only in the strictly one-dimensional case 
one has $\sigma=a_{\perp}$ with $a_{\perp}=\sqrt{\hbar/(m \omega_{\perp})}$ 
the characteristic length of the transverse harmonic confinement.}

By plugging equation \eref{eq:ansz} into the action in equation \eref{eq:az} 
we obtain
\begin{eqnarray}
\fl S[\phi,\sigma] = \int_{t_{1}}^{t_{2}} dt \int dz\ \frac{i\hbar}{2} 
\left( \phi^{\star}\frac{\partial \phi}{\partial t}- \phi \frac{\partial 
\phi^{\star}}{\partial t} \right) -\frac{\hbar^2}{2m} \bigg| \frac{\partial 
\phi}{\partial z} \bigg|^2 
\nonumber 
\\
+ \int_{t_{1}}^{t_{2}} dt \int dz\ - \left[ \frac{\hbar^2}{2m\sigma^2} + 
\frac{1}{2}m\omega_{\perp}^2 \sigma^2 \right]|\phi|^2 - 
\frac{3}{5}\frac{A}{(\sqrt[]{\pi}\sigma)^{\frac{4}{3}}} |\phi|^{\frac{10}{3}}
\end{eqnarray}
{\cblue under the approximation of neglecting the space derivative of $\sigma(z,t)$. 
In the case of weakly-interacting bosons, which are very 
well described in three dimensions by the Gross-Pitaevskii equation  
(i.e. the 3D Schr\"odinger equation with cubic nonlinearity) 
this approximation has been found to be quite good also when 
$\sigma(z,t)$ depends strongly but monotonically on $z$ 
\cite{Salasnich2002,Salasnich2007_2}.} 

Minimizing the action with respect to $\phi$ with the constraint 
\eref{eq:const}, we obtain
\begin{equation}
i\hbar \frac{\partial \phi}{\partial t} = \left[ -\frac{\hbar^2}{2m} 
\frac{\partial^2}{\partial z^2} + \frac{\hbar^2}{2m\sigma^2} + 
\frac{1}{2}m\omega_{\perp}^2 \sigma^2 + 
\frac{A}{(\sqrt[]{\pi}\sigma)^{\frac{4}{3}}} |\phi|^{\frac{4}{3}} - \mu \right] 
\phi \label{eq:phi}
\end{equation}
and by doing the same with respect to $\sigma$
\begin{equation}
\frac{\sigma^4}{a_{\perp}^{4}} = 1 + \frac{4}{5}\frac{A}{\pi^{\frac{2}{3}}}
\frac{m}{\hbar^2}|\phi|^{\frac{4}{3}}\sigma^{\frac{2}{3}}, 
\label{eq:sigma}
\end{equation}
where $a_{\perp}=\sqrt{\hbar/(m\omega_{\perp})}$. We shall refer to 
equation \eref{eq:phi} endowed with \eref{eq:sigma} as the nonpolynomial 
Schr\"{o}dinger equation \cite{Salasnich2002, Salasnich2004, Salasnich2005, 
Salasnich2006, Salasnich2007_1, Salasnich2007_2}.

Let us rewrite the equations in terms of adimensional quantities. 
We can do that with the substitutions
\begin{eqnarray*}
z \rightarrow z a_{\perp} \\ 
t \rightarrow \frac{t}{\omega_{\perp}} \\ 
\sigma(z,t) \rightarrow \sigma(z,t) a_{\perp} \\ 
\phi(z,t) \rightarrow \sqrt[]{\frac{N}{a_{\perp}}} \phi(z,t) \\ 
\mu \rightarrow \hbar \omega_{\perp}(\mu+1) \\ 
A \rightarrow A \frac{\hbar^2}{2m}.
\end{eqnarray*} 
\Eref{eq:phi} then becomes
\begin{equation}
i\frac{\partial \phi}{\partial t} = \left[ -\frac{1}{2} \frac{\partial^2}
{\partial z^2} + \frac{1}{2\sigma^2} + \frac{1}{2} \sigma^2 -1 + \frac{A}{2} 
\frac{N^{\frac{2}{3}}}{\pi^{\frac{2}{3}}}\frac{|\phi|^{\frac{4}{3}}}
{\sigma^{\frac{4}{3}}}  - \mu \right] \phi
\label{eq:phires}
\end{equation}
while equation \eref{eq:sigma} becomes
\begin{equation}
\sigma^4 = 1 + \frac{2}{5}\frac{A}{\pi^{\frac{2}{3}}}N^{\frac{2}{3}}
|\phi|^{\frac{4}{3}}\sigma^{\frac{2}{3}}. 
\label{eq:sigres}
\end{equation} 
{\cblue Remember that in this equation $\sigma$ is adimensional: 
it is in units of the characteristic length $a_{\perp}$ of the transverse harmonic 
confinement. In equation (\ref{eq:sigres}) one can identify two regimes: a regime of strong 
transverse confinement where $\sigma \simeq 1$ and consequently the system is 
truly one-dimensional, but also a regime of weak transverse confinement where $\sigma \gg 1$. 
Thus,} the {\cblue regime of weak transverse confinement} is obtained by neglecting the first 
addend on the right hand side of equation \eref{eq:sigres}, supposing 
that $AN^{\frac{2}{3}}|\phi|^{\frac{4}{3}}\gg 1$. This gives rise to
\begin{equation}
\sigma = \left( \frac{2}{5} A\right)^{\frac{3}{10}} 
\left( \frac{N}{\pi}\right)^{\frac{1}{5}} |\phi|^{\frac{2}{5}},
\label{crucione}
\end{equation}
which plugged back in equation \eref{eq:phires} yields the following 
equation for $\phi$
\begin{equation}
i\frac{\partial \phi}{\partial t} =  \left[ - \frac{1}{2}\frac{\partial^2}
{\partial z^2} +  \frac{1}{2} \left( \frac{5}{2A} \right)^{\frac{3}{5}} 
\left( \frac{\pi}{N}\right)^{\frac{2}{5}} |\phi|^{-\frac{4}{5}} + \frac{7}{10} 
\left(  A \right)^{\frac{3}{5}} 
\left( \frac{5}{2}\frac{N}{\pi}\right)^{\frac{2}{5}} |\phi|^{\frac{4}{5}} - 
\mu -1 \right] \phi.\label{eq:npse}
\end{equation}
In the same approximation we can neglect the term $\propto 
(AN^{\frac{2}{3}}|\phi|^{\frac{4}{3}})^{-\frac{3}{5}}$, and obtain
\begin{equation}
i\frac{\partial \phi}{\partial t} =  \left[ - \frac{1}{2}
\frac{\partial^2}{\partial z^2} + \gamma |\phi|^{\frac{4}{5}} - 
\mu \right] \phi,\label{eq:npse_wc}
\end{equation}
where we defined 
\begin{equation}
\gamma = \frac{7}{10}  A^{\frac{3}{5}} \left( \frac{5}{2}\frac{N}{\pi}\right)^{\frac{2}{5}}.
\label{gammaio}
\end{equation}
Finally, let us express also $\sigma$ in terms of $\gamma$
\begin{equation}
\sigma = \sqrt{\frac{4}{7}}\sqrt{\gamma} |\phi|^{\frac{2}{5}}.
\end{equation}

It will prove useful to study also the limit of strong confinement, 
that is instead given by $\sigma = 1$: 
\begin{equation}
i\frac{\partial \phi}{\partial t} =  \left[ - \frac{1}{2}
\frac{\partial^2}{\partial z^2} + \gamma_{SC} |\phi|^{\frac{4}{3}} - \mu \right] 
\phi.\label{eq:npse_sc}
\end{equation}
\begin{equation}
\gamma_{SC}=\frac{A}{2} \left( \frac{N}{\pi}\right)^{\frac{2}{3}} = 
\frac{5}{4}\left(\frac{4}{7}\gamma \right)^{\frac{5}{3}}.
\end{equation}

\section{Dark solitons in the 1D NPSE}

A soliton is a solution to equation \eref{eq:npse_wc} of the form
\begin{equation}
\phi(z,t) = f(z-vt)e^{i\alpha(z,t)},
\label{eq:phis}
\end{equation}
where $f(\zeta)\geq0$ and $\alpha(z,t)$ are real functions.
Once we substitute \eref{eq:phis} in \eref{eq:npse_wc} we obtain the 
equations for $f$ and $\alpha$
\begin{eqnarray}
f''& = 2\left[ f \left(  \frac{\partial\alpha }{\partial t} + 
\frac{1}{2}\left(\frac{\partial \alpha }{\partial z}\right)^2-\mu \right) 
+ \gamma f^{\frac{9}{5}} \right] \label{eq:f}, \\
f'v& =\left(f' \frac{\partial\alpha }{\partial z} + 
\frac{1}{2}f\frac{\partial^2 \alpha }{\partial z^2}\right) \label{eq:alpha}.
\end{eqnarray}
Upon inspection of equation \eref{eq:alpha}, we find that the right hand 
side should depend only on $\zeta$, so $\frac{\partial\alpha }
{\partial z}$ is only function of $\zeta$.  Let us assume the following 
\begin{equation}
\alpha(z,t) = \theta(z-vt) + \beta(t).
\end{equation}
If we plug this back into equation \eref{eq:f}-\eref{eq:alpha} we get
\begin{eqnarray}
f''& = 2\left[ f \left( \frac{d\beta }{d t} -  v \theta'+ \frac{1}{2}
\left(\theta' \right)^2-\mu \right) + \gamma f^{\frac{9}{5}} \right]  
\label{eq:ft},\\
f'v& =\left(f' \theta' + \frac{1}{2}f\theta''\right). 
\label{eq:faset}
\end{eqnarray}

\subsection{Phase}

Let us solve equation \eref{eq:faset} in order to obtain $\theta'$ 
in terms of $f$. We can rewrite it as
\begin{equation}
v \frac{(f^2)'}{f} =  \frac{(f^2 \theta')'}{f},
\end{equation}
for $f(\zeta) \neq 0$. After integration with respect to $\zeta$ we have
\begin{equation}
f^2(\zeta) \left(v-\theta'(\zeta) \right) = D,
\end{equation}
where $D$ is an integration constant. If we assume that
\begin{eqnarray}
\lim_{\zeta \rightarrow \pm \infty} | f(\zeta) | &=& f_{\infty}, 
\\
\lim_{\zeta \rightarrow \pm \infty}\theta'(\zeta) &=& 0, 
\end{eqnarray}
and we get $D = v f_{\infty}^2$. 
The expression for $\theta'$ we are looking for is then
\begin{equation}
\theta'(\zeta) = v\left(1-\frac{f_{\infty}^2}{f^2(\zeta)} \right),
\label{eq:theta}
\end{equation}
valid for $f(\zeta) \neq 0$.

\subsection{Modulus}

Let us plug equation \eref{eq:theta} back into equation \eref{eq:ft}
\begin{equation}
f'' = 2 \frac{d \beta}{dt} f- v^2 f + v ^2 \frac{f_{\infty}^4}{f^3} - 
2 \mu f + 2 \gamma f^{\frac{9}{5}}.
\label{eq:Newtb}
\end{equation}
We can choose $\beta$ in order to get simpler equations. Let us suppose
\begin{equation}
\beta(t) = \frac{v^2}{2} t,
\end{equation}
so that equation \eref{eq:Newtb} becomes
\begin{equation}
f'' = v^2 \frac{f_{\infty}^4}{f^3} - 2 \mu f + 2 \gamma f^{\frac{9}{5}}.
\label{eq:Newt}
\end{equation}
\Eref{eq:Newt} can be written in the form of a Newton equation for a 
conservative force field
\begin{eqnarray}
f'' &=& -\frac{dW}{df}, 
\\
W &=&  \frac{1}{2} v^2 \frac{f_{\infty}^4}{f^2} +  \mu f^2 - 
\frac{5}{7}\gamma f^\frac{14}{5}.
\end{eqnarray}
The total energy K is then a conserved quantity for all $\zeta$
\begin{equation}
K = \frac{1}{2} f'^2 + W,
\end{equation}
and energy balance gives us the following first order equation
\begin{equation}
K = \frac{1}{2} f'^2 + \frac{1}{2} v^2 \frac{f_{\infty}^4}{f^2} + \mu f^2 
- \frac{5}{7}\gamma f^\frac{14}{5},
\label{eq:energy}
\end{equation}
which is satisfied by the two branches
\begin{equation}
f'= \pm \frac{1}{| f |} \sqrt{2K f^2 - v^2 f_{\infty}^4 - 2 \mu f^4 + 
\frac{10}{7}\gamma f^{\frac{24}{5}}}.
\end{equation}
This equation can be put in the integral form with the separation of 
variables, giving
\begin{equation}
|\zeta - \zeta_{0}| = \int_{f(\zeta_{0})} ^{f(\zeta)} df \frac{f}
{\sqrt{2K f^2 - v^2 f_{\infty}^4 - 2 \mu f^4 + \frac{10}{7}\gamma f^{\frac{24}{5}}}},
\label{eq:int}
\end{equation}
where, since the integral is positive, we have chosen the $+$ sign when 
$\zeta-\zeta_0 \geq 0$ and the $-$ sign otherwise. 
Therefore, $f(\zeta)=f(|\zeta-\zeta_0|)$ has even parity with respect to 
reflections around $\zeta_0$, and we just need to study the case 
$\zeta-\zeta_0 \geq 0$.

We have now to find the values of the parameters $\mu, K, f_{\infty}$. 
This can be done by choosing boundary conditions for $f$.

\subsection{Black solitons}

\begin{figure} []
\centering
\includegraphics[scale=1]{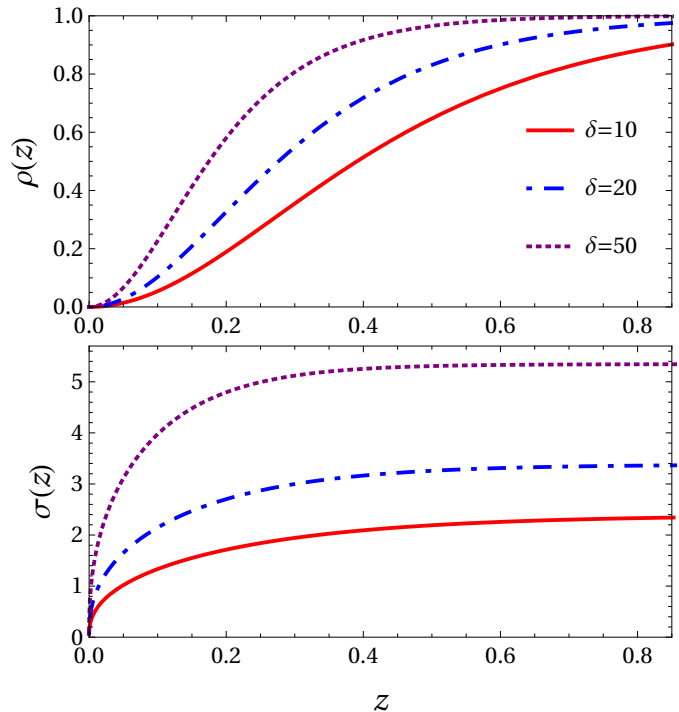}
\caption{\label{fig:bl_} Black soliton. Normalized axial density profile $\rho(z)=h^2(z)
=(f(z)/f_{\infty})^2$ and  
transverse width profile $\sigma(z)$ {\cblue vs axial coordinate $z$, for three 
values of the parameter $\delta=\gamma |f_{\infty}|^{4/5}$, with $\gamma$ 
given by equation (\ref{gammaio}) and $f_{\infty}$ the bulk value of the soliton 
wavefunction.} }
\end{figure}

Let us find an odd parity solution (which features a node in the origin 
for the density $n$) with a positive horizontal asymptote at $+\infty$.
We set $\zeta_0=0$, and since $f$ is even, the phase $\theta$ will account 
for the change of sign. Then, the following boundary conditions are needed
\begin{equation}
f(0)=0,\ \ f(+\infty) = f_{\infty},\ \ f'(+\infty)=f''(+\infty) = 0.
\end{equation}
This is possible only if $v=0$ $(\zeta = z)$ in order to get rid of the 
term in equation \eref{eq:energy} which diverges as $f \rightarrow 0$.
The condition $f''(+\infty) = 0$ applied to equation \eref{eq:Newt} gives
\begin{equation}
\mu = \gamma f_{\infty}^{\frac{4}{5}},
\end{equation}
while $f'(+\infty)=0$ applied to \eref{eq:energy} gives
\begin{equation}
K = \frac{2}{7} \gamma f_{\infty}^{\frac{14}{5}}.
\end{equation}
\Eref{eq:int} then becomes
\begin{equation}
z  = \int_{0} ^{f(z)}  \frac{df}{\sqrt{\frac{4}{7}\gamma f_{\infty}^{\frac{14}{5}}
- 2\gamma f_{\infty}^{\frac{4}{5}}f^2 + \frac{10}{7} \gamma f^{\frac{14}{5}}}}.
\end{equation}
It is possible to get rid of $f_{\infty}$ with the {\cblue rescaling 
\begin{equation}
h(z) = {f(z)\over f_{\infty}} 
\end{equation}
and the parameter 
\begin{equation}
\delta = \gamma f_{\infty}^{4/5} , 
\label{deltaio}
\end{equation}
while } 
$\sigma$ does not change: $\sigma(z) = \sqrt{\frac{4}{7}}\sqrt{\gamma}f(z)^{\frac{2}{5}} 
= \sqrt{\frac{4}{7}}\sqrt{\delta}h(z)^{\frac{2}{5}}$. We obtain
\begin{equation}
z = \frac{1}{\sqrt{\delta}} \int_{0} ^{h(z)}  \frac{dh}{\sqrt{\frac{4}{7} 
- 2h^2 + \frac{10}{7} h^{\frac{14}{5}}}} \label{eq:intnum1}
\end{equation}
Plots of the numerical integration of equation \eref{eq:intnum1} are shown 
in \fref{fig:bl_}, for {\cblue three values of $\delta$.} 
Accordingly with those integrations, also $\sigma(z)$ is computed and 
shown in the plot.

We may define $\tilde{z}=\sqrt{\delta}z$ and $\tilde{\sigma}=\frac{\sigma}
{\sqrt{\delta}}$, so that the theory has no free parameters. This way there 
is only one plot to make, but we want to {\cblue explicitly} show in the plots 
the effects of changing the interaction, so for the moment we are keeping 
$z$ and $\sigma$.

\subsection{Gray solitons}

Let us find an even parity solution with a positive horizontal asymptote. 
We set $\zeta_0=0$ once again, and require the following boundary conditions 
\begin{equation}
f'(0)=0,\ \ f(+\infty) = f_{\infty},\ \ f'(+\infty)=f''(+\infty) = 0.
\end{equation}
The condition $f''(+\infty) = 0$ applied to equation \eref{eq:Newt} gives
\begin{equation}
\mu = \frac{1}{2}v^2 + \gamma f^{\frac{4}{5}}_{\infty},
\end{equation}
$f'(+\infty)=0$ applied to equation \eref{eq:energy} gives
\begin{equation}
K = f^2_{\infty}\left[ v^2 + \frac{2}{7} \gamma f^{\frac{4}{5}}_{\infty} \right],
\end{equation}
while by equating $K$ inside equation \eref{eq:energy} evaluated 
at $0$ and at $\infty$, applying $f'(0)=0$ we get a relation between 
$v, \gamma$ and $f_0$
\begin{equation}
\frac{v^2}{\gamma} = \frac{2}{7}f_0^2\frac{7f^{\frac{4}{5}}_{\infty}f^2_0 
- 5f^{\frac{14}{5}}_{0}- 2f^{\frac{14}{5}}_{\infty}}{2f^2_{\infty}f_0^2- f^4_{\infty}-f_0^4}.
\end{equation}
\Eref{eq:int} then becomes
\begin{equation}
\zeta = \frac{1}{\sqrt[]{\gamma}} \int_{f_0} ^{f(\zeta)} \frac{df\ f}
{\sqrt{\frac{10}{7}f^{\frac{24}{5}}-\frac{v^2}{\gamma}f^4-2 
f^{\frac{4}{5}}_{\infty}f^4+ \frac{v^2}{\gamma}2f^2_{\infty} f^2 + 
\frac{4}{7}f^{\frac{14}{5}}_{\infty}f^2 - \frac{v^2}{\gamma}f^{4}_{\infty}}}.
\end{equation}
Also in this case we can get rid of $f_{\infty}$ in the same way as before, 
obtaining
\begin{eqnarray}
{\cblue \left(\frac{v}{v_s}\right)^2 = \frac{5}{7}h_0^2\frac{5h_0^{\frac{14}{5}}-7h_0^2 +2}
{(1-h_0^2)^2}.}
\\
\zeta = \frac{1}{\sqrt{\delta}}\sqrt{\frac{7}{2}} \int_{h_0} ^{h(\zeta)} 
\frac{dh\ h}{\sqrt{2h^2 -7h^4 + 5h^{\frac{24}{5}}-(2h_0^2-7h_0^4+5h_0^{\frac{24}{5}})
\frac{(1-h^2)^2}{(1-h_0^2)^2}}}
\label{eq:intnum2}
\end{eqnarray}
where of course $h_0 = \frac{f_0}{f_{\infty}}$ {\cblue and  $v_s= \sqrt{\frac{2}{5}\delta}$ is the speed of sound, i.e. the propagation velocity 
of a infinitesimal perturbation in the fluid density.}

Now the theory has two free parameters, namely $h_0$ and $\delta$, 
while $v$ is given in terms of them. In order to be more concise, 
we now define $\tilde{\zeta} = \sqrt{\delta} \zeta$,
$\tilde{\sigma}=\frac{\sigma}{\sqrt{\delta}}$ {\cblue and 
\begin{equation}
\tilde{v}={v\over v_s} , 
\end{equation} 
so} that we are left with a theory that depends only on $h_0$ (or $\tilde{v}$). 
{\cblue The scaled velocity $\tilde{v}$ ranges from $0$ (black soliton) to $1$ (sound wave). 
}

\begin{figure} []
\centering
\includegraphics[scale=1]{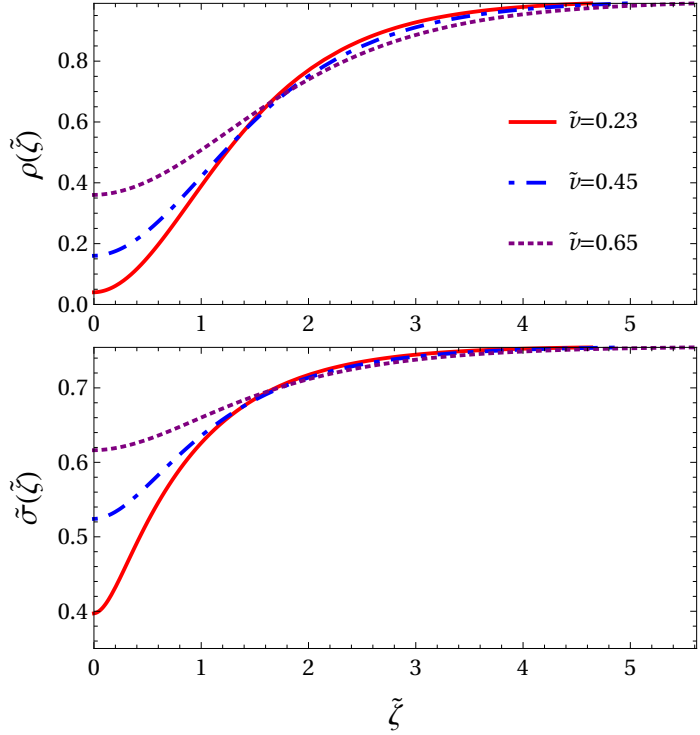}
\caption{\label{fig:gr_} Gray soliton. {\cblue 
Upper panel: Scaled axial density profile $\rho(\tilde{\zeta}) 
= h(\tilde{\zeta})^2=(f(z)/f_{\infty})^2$ vs scaled comoving 
axial coordinate $\tilde{\zeta}=(z-vt)\sqrt{\delta}$. 
Lower panel: Scaled transverse width $\tilde{\sigma}(\tilde{\zeta})$ vs $\tilde{\zeta}$. 
Notice that $\delta=\gamma |f_{\infty}|^{4/5}$, with $\gamma$ 
given by equation (\ref{gammaio}) and $f_{\infty}$ the bulk value of the soliton wavefunction.
$\tilde{v}=v/v_s$ is the velocity rescaled by the sound velocity $v_s$.} 
}
\end{figure}

Plots of the numerical integration of equation \eref{eq:intnum2} are 
shown in \fref{fig:gr_}, for {\cblue three values of $\tilde{v}$.} 
Accordingly with those integrations, also $\tilde{\sigma}(\tilde{\zeta})$ 
is computed and shown in the plot.

\subsection{Phase}

\begin{figure} []
\centering
\includegraphics[scale=1]{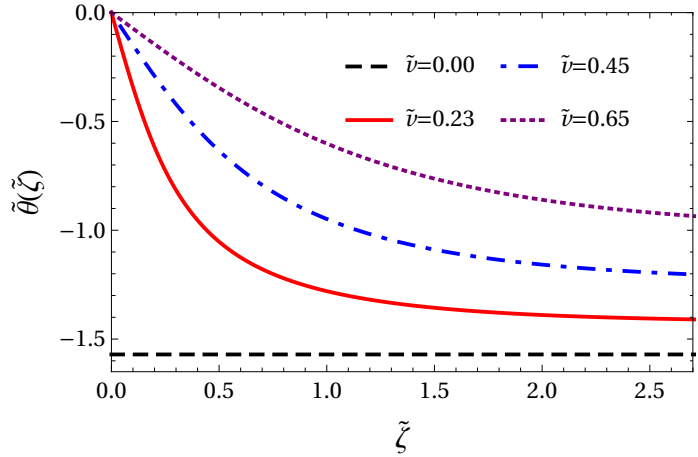}
\caption{\label{fig:th_til_} {\cblue Scaled} phase $\tilde{\theta}(\tilde{\zeta})$ of 
gray ($\tilde{v}>0$) and black ($\tilde{v}=0$) solitons {\cblue vs the scaled 
comoving axial coordinate $\tilde{\zeta}$.}}
\end{figure}

In the case of the black soliton, equation \eref{eq:theta} becomes 
$\theta' = 0$. Since we want the solution to be odd, we choose 
\begin{equation}
\theta(z) =-sgn(z)\frac{\pi}{2} . 
\end{equation}
In the general case, we can now compute $h$ and substitute inside equation 
\eref{eq:theta} to obtain $\theta'$. Let us express also $\theta$ in terms 
of $\tilde{\zeta}$:
\begin{equation}
\theta(\zeta)=\theta(\frac{\tilde{\zeta}}{\sqrt{\delta}})=:\tilde{\theta}
(\tilde{\zeta})
\end{equation}
it follows that
\begin{equation}
\theta'(\zeta)=\sqrt{\delta}\tilde{\theta}'(\tilde{\zeta}),
\end{equation}
therefore
\begin{equation}
{\cblue \tilde{\theta}'(\tilde{\zeta})= \sqrt{\frac{2}{5}}\tilde{v}\bigg(1-\frac{1}{h^2(\tilde{\zeta})}\bigg).}
\label{eq:th_til}
\end{equation}
Last equation upon integration yields $\tilde{\theta}$. A plot of the 
numerical integration of equation \eref{eq:th_til} is shown in \fref{fig:th_til_} for {\cblue four}
values of $\tilde{v}$, {\cblue including} $\tilde{v}=0$, which is the black soliton case.

\section{Weak vs strong transverse confinement}

{\cblue In the previous sections we have mainly investigated the regime of weak transverse 
confinement where $\sigma \gg 1$. See the discussion of equation (\ref{eq:sigres}). 
Let us now briefly analyze some properties of the regime of strong transverse 
confinement where $\sigma \simeq 1$. In this case the system is truly one dimensional 
and the equations are much simpler.} 

By performing computations which are very close to 
the ones we have already shown in the weak confinement case, one can derive 
from \eref{eq:npse_sc} that in
the case of a black soliton the following relation holds
\begin{equation}
z = \frac{1}{\sqrt{\delta_{SC}}} \int_{0} ^{h(z)}  \frac{dh}{\sqrt{\frac{4}{5} 
-  2h^2 + \frac{6}{5} h^{\frac{10}{3}}}},
\end{equation}
where $\delta_{SC}=\gamma_{SC}f_{\infty}^{\frac{4}{3}}$.
Instead for a gray soliton we obtain
\begin{eqnarray}
{\cblue \left(\frac{v}{v_s^{SC}}\right)^2 = \frac{3}{5}h_0^2\frac{3h_0^{\frac{10}{3}}-5h_0^2 +2}
{(1-h_0^2)^2}.}
\\
\zeta = \frac{1}{\sqrt{\delta_{SC}}}\sqrt{\frac{5}{2}} \int_{h_0} ^{h(\zeta)} 
\frac{dh\ h}{\sqrt{2h^2 -5h^4 + 3h^{\frac{16}{3}}-
(2h_0^2-5h_0^4+3h_0^{\frac{16}{3}})\frac{(1-h^2)^2}{(1-h_0^2)^2}}},
\end{eqnarray}

{\cblue where $v_s^{SC}=\sqrt{\frac{2}{3}\delta_{SC}}$. In dimensional units, taking into account equation (\ref{eq:sigma}), 
denoting $\rho_0=\rho(0)$ the minimal axial density of the dark soliton 
and $\rho_{\infty}=\rho(\pm\infty)$ the bulk axial density of the dark soliton, 
one finds that under the condition $\rho_{\infty} a_{\perp} \ll 1$ the dark 
soliton is surely in the strictly 1D regime of strong transverse confinement  
at any point of the axial coordinate. Instead, under the condition 
$\rho_0 a_{\bot} \gg 1$ the dark soliton is surely in the 3D regime of weak 
transverse confinement at any point of the axial coordinate. 
Clearly, in the case of the black soliton, where $\rho_0=0$, 
near the minimum the bosonic cloud cannot be in the regime of weak transverse 
confinement.} 

{\cblue It is important to stress that the comparison of our theoretical results 
with future experiments is constrained not only by the short lifetime of unitary bosonic gas  
due to fast three-body recombinations but also by the lifetime due to the snake 
instability \cite{Antezza2007,Cetoli2013,Lombardi2017,Alphen2018}. 
A dark soliton, that is not strictly 1D, has a snaking transverse 
oscillation which breaks the axial symmetry and eventually destroys the soliton. 
Since the NPSE preserves axial symmetry, it is not suitable for investigating 
snaking oscillations and the dynamics of the dissolving dark soliton. 
However, it is possible to write a triaxial NPSE by using two 
transverse width $\sigma_x(z,t)$ and $\sigma_y(z,t)$ in the ansatz of equation (\ref{eq:ansz}); 
see \cite{Salasnich2009} for details in the case of weakly-interacting bosons.} 

\section{Conclusion}

We have derived the dynamical equations that govern the motion of a unitary 
Bose gas at zero temperature. Using these equations, we have studied dark 
solitons in a unitary Bose gas, confined in a cylindrically symmetric potential.
We have reduced the dimensionality of the 3D nonlinear Schr\"odinger problem 
by means of the 1D nonpolynomial Schr\"{o}dinger equation approach, 
which keeps dynamically into account also the transverse width. We have solved the equations 
employing for the most analytical techniques, in order to find axial density, 
transverse width, phase and velocity of both black and gray solitons, 
in the weak and strong confinement regimes. We have found that the weak 
confinement approximation breaks down in the case of a black soliton. 

Our theoretical predictions could be tested experimentally employing 
Bose gases of alkali-metal atoms at ultra-low temperatures, whose scattering 
length can be tuned by means of Feshbach resonance techniques to reach 
the unitary limit. Despite the difficulties which presently limit the 
time during which such systems can be observed {\cblue due to three-body 
losses but also to the snake instability}, we are confident that the 
recent developments in the experimental techniques are going to allow 
in the future to put our predictions at test.

\end{document}